\documentclass[journal]{IEEEtran}
\ifCLASSINFOpdf
   \usepackage[pdftex]{graphicx}
\else
\fi

\usepackage{amsmath}

\usepackage{ulem}

\begin{document}
%
\title{Photonic spiking neural networks and CMOS-compatible graphene-on-silicon spiking neurons}
%
%
%

\author{Aashu Jha,~ 
        Chaoran Huang,~
        Hsuan-Tung Peng,~
        Bhavin Shastri,~\IEEEmembership{Senior Member,~IEEE,}
        and Paul R. Prucnal,~\IEEEmembership{Life~Fellow,~IEEE}
\thanks{A.Jha, C.Huang, H.T.Peng, P.R.Prucnal are with the Department
of Electrical and Computer Engineering, Princeton University, Princeton, NJ USA (e-mail: aashuj@princeton.edu, chaoranh@princeton.edu, hsuantungp@princeton.edu, prucnal@princeton.edu)}
\thanks{C. Huang is with the Chinese University of Hong Kong, Shatin, Hong Kong (email:crhuang@ee.cuhk.edu.hk)}
\thanks{B. Shastri is with the Department of Physics, Engineering Physics \& Astronomy, Queen's University, Kingston ON, Canada. Also, with the Vector Institute, Toronto ON, Canada.}%
\thanks{A. Jha and C.Huang contribute equally to this work}}

%
%

\markboth{Journal of \LaTeX\ Class Files,~Vol.~14, No.~8, August~2015}%
{Shell \MakeLowercase{\textit{et al.}}: Bare Demo of IEEEtran.cls for IEEE Journals}
%



\maketitle

\begin{abstract}
Spiking neural networks are known to be superior over artificial neural networks for their computational power efficiency and noise robustness. The benefits of spiking coupled with the high-bandwidth and low-latency of photonics can enable highly-efficient, noise-robust, high-speed neural processors. The landscape of photonic spiking neurons consists of an overwhelming majority of excitable lasers and a few demonstrations on nonlinear optical cavities.  The silicon platform is best poised to host a scalable photonic technology given its CMOS-compatibility and low optical loss. Here, we present a survey of existing photonic spiking neurons, and propose a novel spiking neuron based on a hybrid graphene-on-silicon microring cavity. A comparison among a representative sample of photonic spiking devices is also presented. Finally, we discuss methods employed in training spiking neural networks, their challenges as well as the application domain that can be enabled by photonic spiking neural hardware.


\end{abstract}


%
\IEEEpeerreviewmaketitle

\section{Introduction}

Artificial intelligence relies on neural networks, which are getting increasingly computationally intensive and power hungry: the computational power required by neural network models has doubled every 3.4 months in the last decade while the electronic hardware density has doubled every 2 years as per the Moore's law \cite{amodei2018ai}. The need for energy-efficient neural processing has inspired neural hardware engineering, as conventional computers are intrinsically inefficient for distributed processing algorithms \cite{shastri2021photonics}. Building a neural hardware in electronics is bottlenecked by the deceleration of the Moore's law, as well as the fundamental tradeoff between bandwidth and interconnectivity \cite{prucnal2016recent, shastri2021photonics}. This deficiency of electronic hardware has spurred extensive research interest and engendered the nascent field of neuromorphic photonics that aims at leveraging the advantages of optics and neuromorphic architecture to enable a computing platform with high efficiency, interconnectivity and extremely high bandwidth \cite{prucnal2017neuromorphic, shastri2021photonics}.

The vast landscape of neural network models can be broadly divided into artificial neural networks (ANNs), comprised of continuous-valued nonlinear activation functions operating on analog, static inputs, and spiking neural networks (SNNs) which operate on discrete spatiotemporal spikes. The additional temporal dimension gives SNNs a higher representational capacity making them computationally more powerful than their ANN counterparts \cite{maass1997networks, wu2018spatio}. Additionally, the sparsity of input data means the spiking neurons are only active under a spike event, unlike ANNs that continuously process redundant information, which significantly lowers the power consumption of such neural networks \cite{maass1996lower, maass2004computational, amir2017low}. As the demand for more energy-efficient and low-latency neural processing grows, so will the appeal of SNNs. However, there is a need to realize a spiking neural hardware that can actually enable the theoretical power-efficiency and computational superiority of SNNs. Major implementations of spiking hardware in electronics include Loihi from Intel\cite{8259423}, Neurogrid from Stanford \cite{benjamin2014neurogrid}, TrueNorth from IBM \cite{7229264}, etc. Photonics i.e. optical physics offers a viable route to further push the energy efficiency and processing speed of SNNs. 

Existing photonic spiking hardware are overwhelmingly based on excitable semiconductor lasers on III-V platforms \cite{shastri2016spike, nahmias2013leaky, peng2019temporal, hurtado2012investigation, barbay2011excitability}, which undergo high optical loss due to weak confinement of light and high material absorption, and can be prohibitive to scaling. Such scalability issues can be resolved by a silicon photonic platform, which is CMOS-compatible and benefits from existing fabrication technology, and have very low loss at telecom wavelengths. There has been some recent work in realizing spiking neuron functionality on silicon; most notably Refs. \cite{chakraborty2018toward, feldmann2019all} use phase-change materials embedded on silicon and silicon nitride platforms to engineer spiking-like functionality through nonlinear pulse transformation. However, the spiking behavior in their approaches is reliant on a synchronized operation between the output spike pulses and the input data. Their approaches face two fundamental limitations: the lack of temporal encoding feature, which is a key characteristic of spiking, and the lack of asynchronicity prohibits arbitrary network configurations. In this paper, we propose a CMOS-compatible spiking neuron based on a graphene-on-silicon nonlinear microring resonator where the nonlinear effects in silicon and graphene generates a spiking dynamical system. Graphene enhances the efficiency of the nonlinear photonic processes \cite{ishizawa2017optical, gu2012regenerative, feng2019enhanced} that enables spiking at a much faster timescale ($\approx$picoseconds) previously impossible with silicon-only nonlinear devices \cite{van2012cascadable, yacomotti2006fast}. Additionally, realizing spiking on a standard commercially-available silicon platform offers the advantages of large-scale manufacturing and easier interface with standard silicon photonic components enabling it as a practical technology.         

Another crucial component of realizing an efficient neural network besides hardware is algorithms. ANNs have gathered mass acceptance due to their simplicity in training and availability of large labelled datasets. ANNs can learn on gradient-descent based algorithms like backpropagation \cite{726791}, whereas SNNs require specialized algorithms due to the non-differentiable nature of spiking \cite{roy2019towards, pfeiffer2018deep}. The constraints become further stringent when it comes to a photonic hardware, which is highly susceptible to process variations and noise, and necessitates algorithm-hardware co-design. While training algorithms for spiking can be adapted from neuroscience and electronics, it will be key to be mindful of the requirements of photonic hardware and the application when selecting an algorithm.   

Once there is a framework for spiking neural hardware and algorithms in place, it is also advantageous to evaluate the spike-based processing on event-based applications that naturally fall within the realms of spike-based processing. This goes against the approach that has been taken so far of chasing ANNs over classification accuracy. Event-based applications like brain-machine interfaces, autonomous driving etc, that also demand ultra-fast and low-power computation and are beyond the grasp of ANNs and spiking electronics are where we should deploy photonic spiking hardware. Availability of large dynamic datasets of such applications will be useful for efficient training before deployment.

In this paper, we provide a survey of existing demonstrations of photonic spiking neurons, propose a novel CMOS-compatible spiking neuron design based on a graphene-on-silicon microring resonator and compare a representative sample of neurons against various performance metrics. We then summarize various training methods available to spiking neurons, show a simulated benchmark classification on our proposed neuron using one such class of algorithm and argue the case for local learning algorithms for photonic hardware. We finally discuss the application domain of spiking neural hardware that can best showcase their computational strength.


\section{Overview of photonic spiking neurons}
Existing demonstrations of photonic spiking neurons mirror biological neurons the closest in behavior rather than emulating them. Research in neuroscience has led to various models of biological neurons, encompassing a wide range of complexity from the earliest ones like the Hodgkin-Huxley model, which was computationally complex, to progressively simplified ones like the Izhikevich neuron model and then the leaky integrate-and-fire model. Developing a neural hardware entails replicating a given model's behavior. Laser cavities have long been studied for excitable and spiking properties \cite{krauskopf2003excitability, mohrle1992gigahertz, dubbeldam1999excitability, dubbeldam1999self, wunsche2001excitability, pan1993optical}. However, the work of Nahmias et al.\cite{nahmias2013leaky} drawing parallel between semiconductor lasers and the leaky \textit{integrate-and-fire} neuron model rekindled interest in photonic spiking hardware. Consequently, an overwhelming majority of photonic spiking neurons are based on semiconductor lasers, while the remainder are variants of nonlinear optical cavities. One way to systematically categorize them can be to distinguish depending on whether the operation mechanism is opto-electronic or all-optical. Figure \ref{fig:overview} illustrates the categories, including a cartoon illustration of a representative system in each sub-category.     

\subsection{Opto-electronic systems}
\begin{figure}
    \centering
    \includegraphics[width=0.4\textwidth]{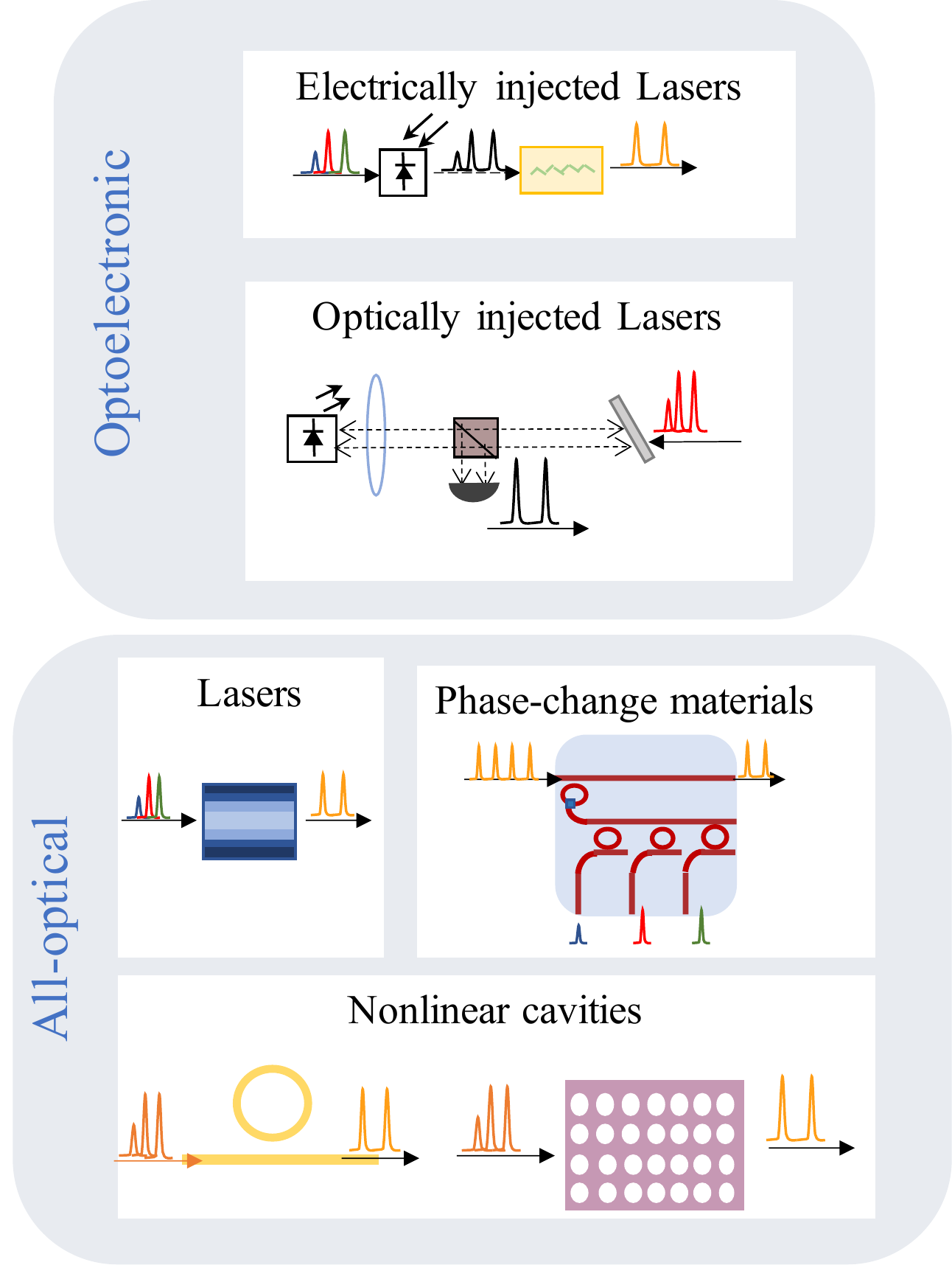}
    \caption{Implementations of photonic spiking neurons, categorized into (top) optoelectronic devices including semiconductor lasers, and (bottom) all-optical devices, including lasers, phase-change material-based devices and nonlinear cavities. Additionally, the operational principles of each class of devices is also illustrated, showing incoming and output spikes, where colors represents optical wavelengths.}
    \label{fig:overview}
\end{figure}

This class of spiking systems includes semiconductor lasers where optical feedback within the laser cavities leads to nonlinearities that endow excitability. Refs. \cite{pengneuro, prucnal2016recent} offer  an exhaustive overview of the spiking neurons that fall under this class. The lasing mechanism is based on electrical pumping while the injection can be either electrical or optical. The electrically pumped devices are often studied with the Yamada model, which was first used to demonstrate excitability in lasers with saturable absorbers \cite{PhysRevE.60.6580}. Notable examples within this include: graphene excitable laser \cite{shastri2014simulations}, VCSELs \cite{nahmias2013leaky}, semiconductor optical amplifiers \cite{PhysRevE.68.036209} and  distributed feedback laser \cite{pengneuro}. An illustrative operating principle of a distributed feedack laser is shown in Fig. \ref{fig:overview} where the weighted optical signal is summed by a photodetector which drives the excitable laser resulting in output spikes. 

Another class of opto-electronic lasers are optically injected where the optical signal directly drives the excitable laser resulting in spike response, as shown in Fig. \ref{fig:overview}. Early demonstration of excitability in lasers subject to optical injection was shown in \cite{PhysRevLett.88.063901, PhysRevLett.88.023901}. This was followed by excitable devices involving quantum dots \cite{PhysRevLett.98.153903}, microring lasers \cite{PhysRevA.81.033802, gelens2010excitability}, microdisk lasers \cite{alexander2013excitability} and quantum wells \cite{kelleher2011excitability}. 

\subsection{All-optical systems}
All-optical excitable devices have relatively more diversity in their operational principles, ranging from lasers to passive optical cavities to exotic material-enhanced cavities, as shown in Figure \ref{fig:overview}.  
Lasers within this category include those that are optically pumped and become excitable in response to optical perturbations. Early works of this kind were done in Q-switched lasers \cite{Spuhler:99} and lasers with saturable absorber \cite{larotonda2002experimental}, \cite{selmi2014relative}.  This was followed by  several implementations of vertical cavity surface emitting lasers (VCSELs) \cite{barbay2011excitability, hurtado2012investigation}. There was a parallel effort in studying all-optical excitability in optical cavities such as microring resonators \cite{van2012cascadable} and photonic crystal cavities \cite{yacomotti2006fast}. Recently a lot of attention is drawn towards realizing engineered excitability using phase change materials \cite{feldmann2019all, chakraborty2018toward}.   \\  

Laser-based systems face the issue of scalability as their epitaxial and structural characteristics require non-standard specialized fabrication methods. On the other hand, integrated devices that can be fabricated at scale in a commercial foundry process has significantly higher odds of becoming a viable technology. Additionally, photonic fabrication technology is highly susceptible to process variation, and almost always requires electronics-assisted post fabrication compensation e.g. tuning resonator resonance wavelengths. This need for co-existing photonics and electronics requires photonics to remain compatible to CMOS electronics. These reasons together advocate for spiking neurons on CMOS compatible platforms like silicon. Prior demonstrations of a CMOS-compatible spiking neuron either suffer from speed bottlenecks, or lack the asynchronicity fundamental to spiking neurons. In the following section, we propose an alternative approach of engineering a CMOS-compatible spiking neuron.

\section{A CMOS-compatible spiking neuron}
In this section, we first present our nonlinear coupled-mode theory based model incorporating the nonlinear effects in silicon and graphene, and the basis of excitability in the microring. Using this model, then we show the simulation results showing key characteristics of a spiking neuron in a graphene embedded silicon microring cavity. These include: asynchronous spike generation in response to input perturbation, threshold operation, temporal integration, and cascadability. 

Excitability, and consequently spiking, in our device arises from silicon's nonlinear optical effects \cite{lin2007nonlinear}. Similar approaches have been undertaken before: Ref. \cite{van2012cascadable} demonstrated excitability through competing thermal and free-carrier dynamics, but the processing speed was capped to the MHz range due to the slower timescale of thermal effects. For excitability at higher speeds, faster mechanisms such as free-carrier and instantaneous Kerr effects can be used. However, such a fast excitable system has not yet been experimentally demonstrated due to the high optical power threshold to enter such a nonlinear regime \cite{van2016optical}. We address this issue by incorporating graphene in the nonlinear dynamical system. Our proposed spiking neuron is based on a hybrid graphene-embedded silicon microring resonator (MRR). Graphene has previously been utilized for enhancing efficiency in several nonlinear silicon devices \cite{ishizawa2017optical, gu2012regenerative, feng2019enhanced}. In \cite{gu2012regenerative}, the authors even show a power threshold reduction of 50 times in a graphene-si cavity over just monolithic silicon. The combination of the MRR cavity-induced coherent power buildup and the enhancement of silicon nonlinearities by graphene improves the overall power efficiency of the fast excitability as compared to just silicon.

\subsection{Coupled mode theory model for the hybrid microring}

\begin{figure}[htbp]
    \centering
    \includegraphics[width=0.5\textwidth]{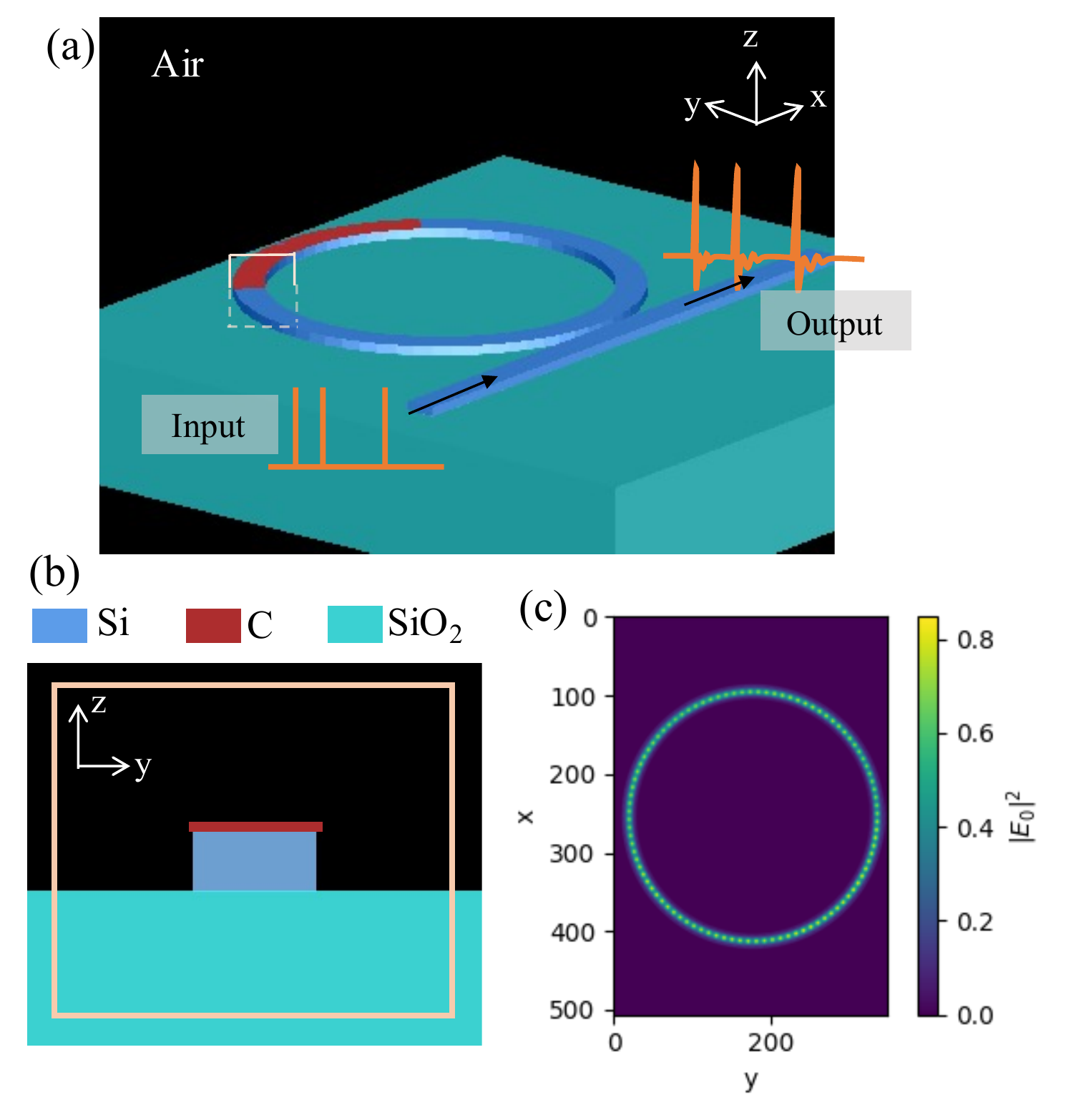}
    \caption{(a) Perspective view of the hybrid graphene-on-silicon microring coupled with a silicon bus waveguide, (b) YZ cross-section of the microring waveguide showing graphene overlaid on silicon, (c) Normalized electric field intensity, $|E_0|^2$ profile of the resonant TE mode at a midsection of the waveguide, as calculated by 3D FDTD solver on Lumerical.}
    \label{fig:design}
\end{figure}

A variety of photonic nonlinear effects are at play in a hybrid silicon-graphene microring cavity which together form the basis of the spiking dynamical system. A good introduction to silicon photonic nonlinearities can be found in \cite{lin2007nonlinear}, and the major effects in integrated silicon waveguides are shown in Fig. 1 of Ref. \cite{jha2020high}. Due to the centrosymmetry in silicon crystalline structure, the nonlinear effects of interest are the first- and the third- order effects. The third-order nonlinear effects are instantaneous parametric processes and manifest as intensity-dependent dispersion, known as the Kerr effect, and absorption called two-photon absorption (TPA). TPA generates free-carriers that trigger first-order nonlinear effects namely, free carrier absorption (FCA) and dispersion (FCD). The free-carrier effects tend to be much stronger than the third-order effects in integrated SOI \cite{lin2007nonlinear}. All the absorption mechanisms contribute to thermal nonlinearity, which is a slow ($\mu$s) dispersive effect and can be assumed to be constant for fast (GHz) signals. Kerr and thermal effects cause a red-shift in the cavity resonance while FCD causes a blue-shift. This is why FCD and Kerr effects are generally considered to be competing effects in nonlinear signal processing. Graphene also exhibits the third-order Kerr and TPA effects, and a strong first-order absorptive effect, known as saturable absorption. There are two main reasons as to why graphene can enhance efficiency of nonlinear silicon photonic effects. First, the Kerr effect in graphene is much higher than the silicon Kerr effect, and it acts in the same direction as the silicon FCD effect \cite{ataloglou2018nonlinear}. Thus the Kerr effect in graphene amplifies the FCD effect in silicon. Second, the saturable absorption in graphene induces a nonlinear dependence of the cavity quality factor on intensity, which facilitates nonlinear effects and reduces the power threshold for bistability \cite{ataloglou2018nonlinear}. 


To trigger the aforementioned nonlinearities, a microring resonator (MRR) is used to allow for coherent optical intensity buildup. A perspective view of the graphene-on-silicon MRR and a waveguide crosssection are studied in this work is shown in Figs.\ref{fig:design}(a) and (b). The dimensions of the microring are given in Table \ref{params}. The high refractive index of silicon allows for strong light confinement to facilitate light-matter interaction in both silicon and graphene. Using a 3D finite-difference time domain (FDTD) simulation on Lumerical, the electric field distribution corresponding to the resonant TE mode in the cavity was calculated. Electric field intensity at the waveguide  mid-section is shown in Fig.\ref{fig:design}(c). We model light propagation in the microring using a nonlinear coupled-mode theory approach based on \cite{chen2012bistability}, with the inclusion of graphene contributions, i.e. its Kerr effect and saturable absorption in our updated model. The coupled ordinary differential equations for the temporal evolution of normalized cavity light amplitude $a$ and free-carrier density $n$ are given in Eq. \ref{dadt} and \ref{dndt}:

\begin{align}
    \delta a/\delta t & =\sqrt{P} + i(\delta a - n_{\text{kerr}}|a|^2a)
    +i(n+\sigma_{\text{FCD}}n^{0.8})a \\
    &- (1+\gamma_{\text{FCA}}n)a  
    -\alpha_{\text{TPA}}|a|^2a - (\frac{1}{1+\frac{|a|^2}{W_\text{sat}}})a
\label{dadt}  
\end{align}

\begin{equation}
    \delta n/\delta t = -\frac{n}{\tau} +|a|^4 
\label{dndt}    
\end{equation}

The time variable $t$ is normalized with respect to $\frac{1}{\Gamma_0}$ where $\Gamma_0 = \omega_0/2Q_L$ where $\omega_0$ is the microring resonance frequency, and $Q_L$ is the loaded quality factor, which includes contributions from microring radiation loss, coupling loss and saturable absorption loss in the low-intensity limit in graphene. The cavity energy variable is normalized as $|a|^2 = |u|^2\sqrt{\sigma \beta}$, where $u$ is the unnormalized cavity mode energy and $\sigma = \sigma_{r1}\frac{\omega_0}{n_0\Gamma_0}$ and $\beta = \frac{c^2\beta_{2,\text{Si}}}{\Gamma_02\hbar\omega_0n_0^2V_{\text{TPA}}V_{\text{car}}}.$ where $\beta_{2,\text{Si}}$ is the TPA coefficient of silicon. $V_{\text{TPA}}$ is the nonlinear TPA volume parameter defined as per Ref.\cite{Uesugi:06}; it quantifies the overlap of light and the nonlinear silicon material, and was calculated using Eq. \ref{vtpa}:

\begin{equation}
    V_{\text{TPA}} = \frac{\int \beta(r)_{\text{2}}n(r)^2|E_0(r)|^4 d^3r}{(\int n(r)^2|E_0(r)|^2d^3r)^2 \beta_{2}}
    \label{vtpa}
\end{equation}
Here $\beta(r)$, $n(r)$ and $E_0(r)$ are spatially dependent variables. $E_0(r)$ and $n(r)$ were obtained from the FDTD simulation for the resonant fundamental TE mode where the integration volume is defined over a 3d grid overlaying the microring. We set $\beta(r) = \beta_{2,\text{Si}}$ for when $n = 3.48$, i.e. within the silicon waveguide. A two-dimensional crosssection of the resonant mode profile is shown in Fig. \ref{fig:design}(c). Similarly, $V_{\text{car}}$ is the nonlinear carrier volume, which was calculated as $V_{\text{car}} = 2\pi RL_D$, where $R$ is the microring radius and $L_D$ is the carrier diffusion length which is approximate as the waveguide width, similarly as Ref. \cite{Uesugi:06}.  

In the first equation, $P$ corresponds to the input light power, and is normalized as $P = \frac{\sigma\beta \Gamma_c}{\Gamma_0}P_{\text{in}}$ where $P_{\text{in}}$ is the input light power and $\Gamma_c$ is the coupling coefficient between the bus waveguide and the microring. The imaginary terms on the right hand side relate to the linear and nonlinear dispersive effects such as the Kerr and FCD effects. $\delta$ is the normalized frequency detuning between microring resonance frequency and the input light wavelength. $n_{\text{kerr}}$ contains the Kerr contributions of both silicon and graphene, and is defined in Eq. \ref{nkerr}:
\begin{equation}
    n_{\text{kerr}} = \frac{1}{\sqrt{\sigma\beta}\Gamma_0}(\gamma^{\text{Si}}_{\text{kerr}} + \gamma^{\text{G}}_{\text{kerr}})
    \label{nkerr}
\end{equation}
where the $\gamma^{\text{Si}}_{\text{kerr}}, \gamma^{\text{G}}_{\text{kerr}}$ parameters are defined as per Eqs. \ref{gamma1}, \ref{gamma}:
\begin{equation}
    \gamma^{\text{Si}}_{\text{kerr}} = \frac{\omega_0cn^{\text{max}}_{2,\text{Si}}}{V^{\text{Si}}_{\text{kerr}}} \\
    \label{gamma1}
\end{equation}
\begin{equation}
    \gamma^{\text{G}}_{\text{kerr}} = \frac{\sigma^{\text{max}}_{3,\text{Im}}}{V^{\text{G}}_{\text{kerr}}\epsilon_0^2}
\label{gamma}
\end{equation}    

$n^{\text{max}}_{2,\text{Si}}$ is the Kerr coefficient of silicon and $\sigma^{\text{max}}_{3,\text{Im}}$ is the third-order nonlinear conductivity of graphene which  quantify the strength of the Kerr nonlinearity in each material.  $\sigma^{\text{max}}_{3,\text{Im}}$ is calculated using $n^{\text{max}}_{2,\text{G}}$ for a graphene layer thickness of 0.33 nm \cite{vermeulen2016negative, chatzidimitriou2015rigorous}.  $V^{\text{G}}_{\text{kerr}}$ and $V^{\text{Si}}_{\text{kerr}}$ are nonlinear Kerr volumes which measure the overlap between light and silicon and graphene respectively. They are defined as per the definition in Ref. \cite{Uesugi:06}, given in Eq. \ref{vkerrsi}, \ref{vkerrg} below: 
\begin{equation}
    V^{\text{Si}}_{\text{kerr}} = \frac{\int n_{\text{2}}(r)n(r)^2|E_0(r)|^4 d^3r}{(\int n(r)^2|E_0(r)|^2d^3r)^2 n_{2}}
    \label{vkerrsi}
\end{equation}
\begin{equation}
    V^{\text{G}}_{\text{kerr}} = \frac{\int\sigma_{3,\text{Im}}(r)|E_{0,\parallel}(r)|^4 d^3r}{(\int n(r)^2|E_0(r)|^2d^3r)^2\sigma^{\text{max}}_{3,\text{Im}}}
    \label{vkerrg}
\end{equation}
$V^{\text{Si}}_{\text{kerr}} = V_{\text{TPA}}$ as the nonlinear volumes primarily rely on the confinement of light in the nonlinear material rather than the magnitude of the nonlinear parameter itself.

The second term in the RHS of Eq. \ref{dadt} corresponds to the free-carrier dispersion, where the $\sigma_{\text{FCD}}$ term is defined as per Ref. \cite{chen2012bistability}, i.e. $\sigma_{\text{FCD}} = \frac{\omega_0\sigma_{r2}}{n\Gamma_0\sigma^{0.8}}$. The real parts of the RHS of Eq. \ref{dadt} encompass the loss mechanisms: linear loss, free-carrier dependent loss ($\gamma_{\text{FCA}}$) and two-photon absorption $\alpha_{\text{TPA}}$, which are defined as Eq. \ref{atpa}, \ref{yfca}:
\begin{equation}
   \alpha_{\text{TPA}} = \frac{\beta_2c^2}{2n^2\Gamma_0V_{\text{TPA}}\sqrt{\sigma\beta}}
   \label{atpa}
\end{equation}

\begin{equation}
    \gamma_{\text{FCA}} = \frac{\sigma_{\text{FCA}}c}{2N\Gamma_0\sigma}
    \label{yfca}
\end{equation}

The last term in Eq. \ref{dadt} accounts for the intensity-dependent saturable absorption in the graphene layer, which is derived from the saturable absorption lifetime model given in Ref. \cite{ataloglou2018nonlinear}. $W_{\text{sat}}$ is the total stored energy in the cavity at the onset of saturable absorption, and is calculated as $W_{\text{sat}} = (E_{\text{sat}}/E_{\parallel})^2$ where $E_{\parallel}$ is the normalized electric field component tangential to the graphene layer and $E_{\text{sat}} = \sqrt{2\eta_0 I_{\text{sat}}}$ where $I_{\text{sat}}$ is the saturation intensity. 

Eq. \ref{dndt} models the temporal evolution of normalized carrier density in the microring. $\tau = \tau_{\text{car}}\Gamma_0$ is the normalized free-carrier lifetime in the cavity. Its dependence on $|a|^4$ alludes to the TPA-mediated generation of free-carriers. Therefore our  model in Eq.(1) and (2) encompasses all the nonlinear effects that occur in the hybrid graphene on silicon microring cavity. The parameter values used in our simulation are tabulated in Table \ref{params}. 

\begin{table}[]
\caption{Simulation parameters used in this work.}
\begin{tabular}{cccc}
Parameter           & Value                               & Parameter & Value \\  \hline
$n_{2, \text{Si}}$   & $4.5\times10^{-18}$ Wm$^{-2}$ \cite{bristow2007two} &       R    &     5e-6   \\
$n_{2, \text{G}}$ & $-1\times10^{-13}$ Wm$^{-2}$ \cite{dremetsika2016measuring} &        $ W_\text{Si}$  & 500 nm       \\
$\beta_{2, \text{Si}}$            & $9\times 10^{-8}$ \cite{bristow2007two}          &   $H_\text{Si}$  & 220 nm     \\
$\tau_{\text{SA}}$   &     20 ps                                 & $\tau_{\text{car}}$          &     20 ps   \\
$\sigma_{FCA}$       &  $1.45\times 10^{-23}$   \cite{chen2012bistability}&                               $Q_0$                &   60e3                            \\
$I_{\text{sat}}$     &  $10^{10}$ \cite{marini2017theory}   &                            $Q_e$                &   10e3                                   \\

$\sigma_{r1}$        & 8.8e-28 m$^3$ \cite{soref1987electrooptical}                        &    $n_{\text{Si}}$  &   3.478     \\
$\sigma_{r2}$        & 1.35e-22 m$^3$ \cite{soref1987electrooptical}                   &      $n_{\text{SiO2}}$ &  1.44    \\\hline 
\end{tabular}
\label{params}
\end{table}

\begin{figure}[h!]
    \centering
    \includegraphics[width=0.5\textwidth]{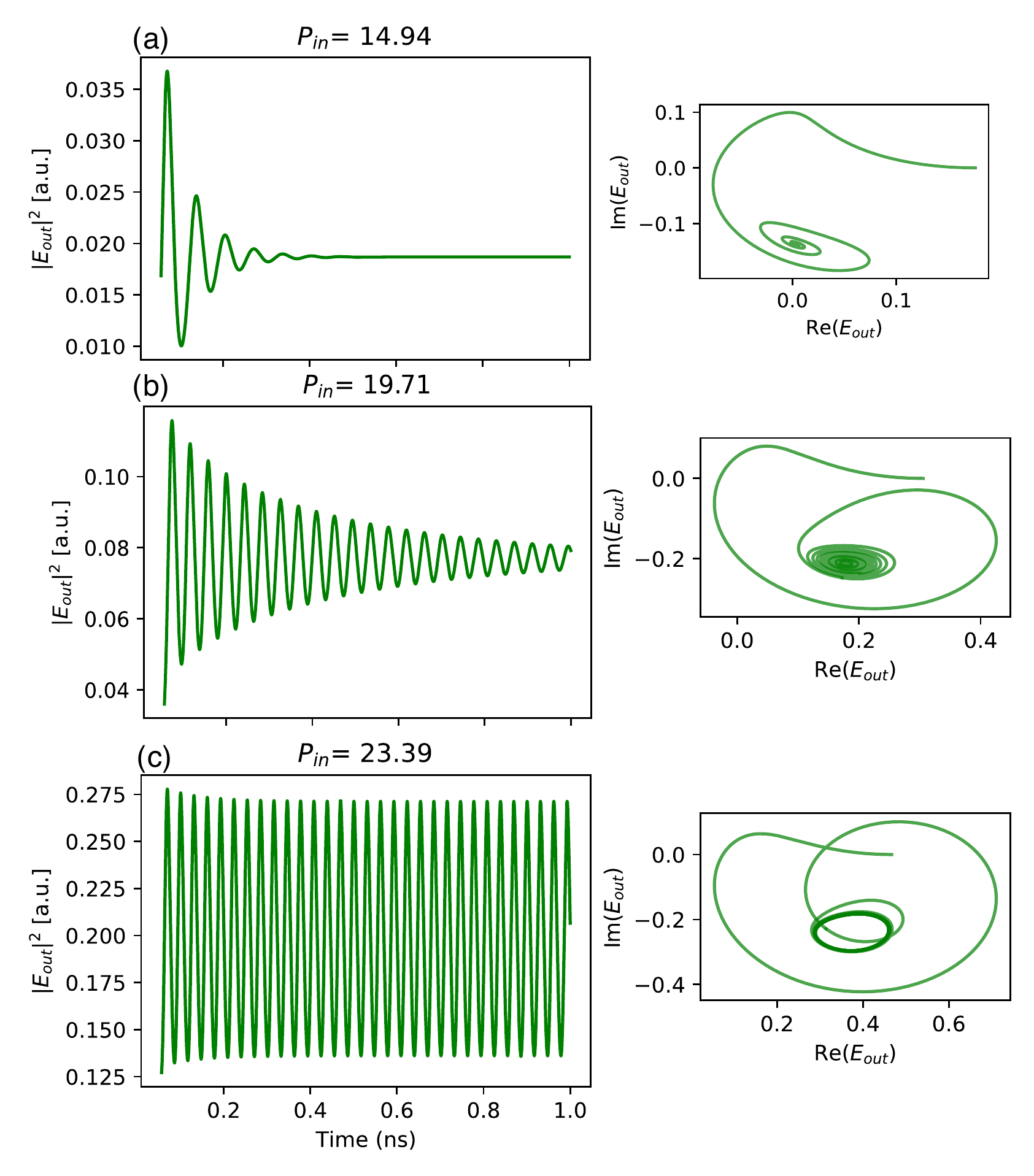}
    \caption{Dynamics of characteristic class II excitability. (a)-(c) show the temporal evolution of output power and (i)-(iii) show the corresponding phase space portraits of cavity light amplitude (u), in response to static input power to the MRR. The input power increases from top to bottom. The titles of (a)-(c) note the corresponding simulation parameters (normalized input power ($P_{IN}$), normalized frequency detuning ($\delta$), saturable absorption lifetime ($\tau_{SA}$) and free-carrier lifetime ($\tau_c$) in picoseconds.}
    \label{sp}
\end{figure}

\subsection{Analogy of the CMT model to the spiking neuron model}
The simplest spiking neuron model that emulates the biological neuron behavior is the \textit{integrate-and-fire} model. Its computational efficiency has made it the most popular model studied in computational neuroscience. Eq. \ref{if} represents a simplified version of this model, where for the $i$th neuron of the network, $y_i$ is the neuron state variable, $a_i$ corresponds to the resting state, $b_i$ is the internal neuron parameter, $\delta$ corresponds to the spikes from connecting neurons with $s_{ij}$ representing the synaptic connections.
\begin{equation}
    \dot{y_i} = a_i + b_iy_i +\sum_{j=1}^n s_{ij} \delta(t-t_j^*)
    \label{if}
\end{equation}
An analogue to the \textit{integrate-and-fire} model is the \textit{resonate-and-fire} model proposed in Ref. \cite{izhikevich2001resonate}. Eq \ref{rf} describes this neuron model, which closely resembles Eq. \ref{if}. Here, $z_i$ is the neuron state variable, $\delta$ corresponds to the spikes from connecting neurons with $c_{ij}$ representing the synaptic connections akin to the $y_i$, $\delta$ and $s_{ij}$ terms in the \textit{integrate-and-fire} model. The only difference is in the internal neuron parameter, $(i\omega_i + b_i)$, which has an imaginary component. This imaginary parameter introduces oscillatory dynamics in the neuron that is absent in \textit{integrate-and-fire} neurons.  

\begin{equation}
\dot{z_i} = (i\omega_i + b_i)z_i +\sum_{j=1}^n c_{ij} \delta(t-t_j^*)
    \label{rf}
\end{equation}

Now we draw the analogy between our coupled-mode theory model of the microring and the \textit{resonate-and-fire} model described in Eq.\ref{rf}. To this end, we simplify Eq. \ref{dadt} which describes the evolution of the neuron state variable, $a$ i.e. the light amplitude in the microring. Eq. \ref{simp} is a simplified version of Eq. \ref{dadt}. 

\begin{equation}
    \dot{a_i} = (i\Theta_i + B_i)a_i + I_i
    \label{simp}
\end{equation}
The parameters introduced above are defined as follows:\\
$I_i = \sqrt{P},$ corresponds to the input, $B_i= 1+ \gamma_{\text{FCA}}n-\alpha_{\text{TPA}}|a|^2- \frac{1}{1+\frac{|a|^2}{W_\text{sat}}},$ represents the amplitude decay mechanisms and 
$\Theta_i = \delta-n_\text{kerr}|a|^2+n +\sigma_{\text{FCD}}N^{0.8}$ represents the dispersive contributions. Comparing Eqs. \ref{rf} and \ref{simp} clarifies the analogy between the two models. $I_i$ is equivalent to the cumulative input term $\sum_{j=1}^Nc_{ij}\delta(t-t_j^*)$. Similarly,    $(i\Theta_i + B_i)$ is functionally equivalent to the internal neuron parameter term, $(i\omega_i+b_i)$. As we investigate the dynamics of our neuron model in the following subsections, we will witness the oscillatory dynamics introduced in the system by the $\Theta$ term, as is characteristic of \textit{resonate-and-fire} neurons.         

\subsection{Steady-state characteristics}
The first step to emulate a spiking neuron in the microring is to show excitable dynamics. Excitability occurs when a perturbation from a system's rest state results in a large excursion of physical variables, i.e. output light in our case, followed by an eventual rest back to the equilibrium. In \cite{van2012cascadable}, the authors show the existence of a class II excitable system in a silicon microdisk close to a regime of self-pulsation. Self-pulsation refers to the phenomenon  where light oscillates between two output states in response to a constant input light power. In addition to Ref. \cite{van2012cascadable}, it has also been previously reported in several other photonic cavities \cite{brunstein2012excitability, chen2012bistability}.

Motivated by their results, we look for a self-pulsation regime in our device. To do that, we characterize the temporal dynamics of output light in response to a continuous wave (CW) optical input by solving the ODEs in Eq. \ref{dadt}, \ref{dndt}.  Fig. \ref{sp} shows the temporal evolution of output power, $|E_{\text{out}}|^2$ in response to different input light powers, $P_{\text{in}}$, as well as the corresponding phase portraits of the output light amplitude, $E_{\text{out}}$. Fig. \ref{sp}(a) shows that in a relatively low input power regime, the output light amplitude decays to a steady state and the corresponding phase portrait in the right plot shows decay to a stable fixed point. By increasing the input power further (Fig. \ref{sp}(b)), the output undergoes damped oscillatory decay to a steady state, where the phase portrait reveals oscillations and slow decay to a fixed point. Finally, a relatively higher input power (Fig. \ref{sp}(c)) results in a stable oscillating output, while in the phase portrait, a stable limit cycle is formed. This is the regime of self-pulsation, which results from the interaction between FCD-induced blue-shift of the microring resonance frequency and the light intensity in the MRR, much like in \cite{chen2012bistability}. The only difference is that in our case, graphene Kerr boosts the FCD effect, and the saturable absorption lifetime, which determines the photon cavity lifetime,  ensures that the photon cavity lifetime is roughly the same order of magnitude as the free-carrier lifetime.
Such a transition from decay to oscillatory decay to a stable oscillatory state is characteristic of a class II excitable dynamical system. Additionally, the oscillatory decay to steady state is indicative of a \textit{resonate-and-fire} neuron behavior as discussed in the previous section.  



\subsection{Spiking neuron dynamics}

\begin{figure}[h!]
        \centering
        \includegraphics[width=0.4\textwidth]{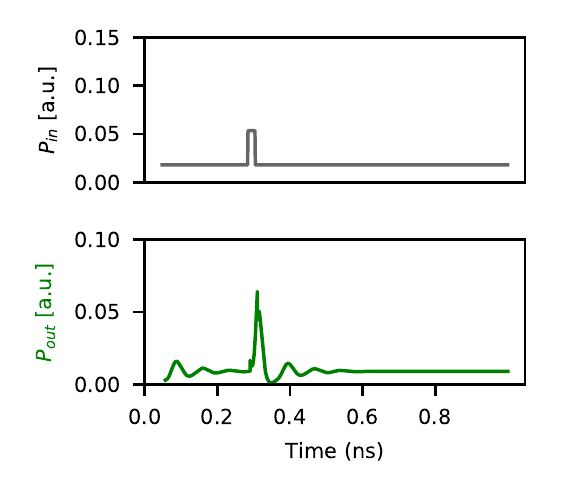}
        \caption{Spiking behavior in the MRR. (top) Input light with perturbation, and (bottom) corresponding output with a spike response to input perturbation}
        \label{fig:spike}
    \end{figure}

Excitability in a spiking system is characterized by the following properties: 1. temporal integration of incoming pulses, 2. generation of a spike for inputs above a threshold, and 3. asynchronous spike generation without being triggered on input. These properties enable the repeatable nature of a spiking dynamical system that assures computational power efficiency and noise robustness. We shall investigate whether these properties exist in the hybrid microring system. 
Similar to \cite{van2012cascadable}, we look for excitability in proximity to the self-pulsating regime.  More specifically, excitability in this case means for a constant input power slightly below the self-pulsation threshold, if there is a perturbation that crosses the self-pulsation threshold, it can throw the system into a limit cycle-like state resulting in an output spike. Fig. \ref{fig:spike} shows a perturbed input, where the constant power level is below while the perturbation is above the self-pulsation threshold. The input perturbation results in an output spike. This illustrates the excitable property of the microring that results in spiking. Unlike the typical integrate-and-fire neurons \cite{peng2019temporal, shastri2016spike} where upon stimulation, the output undergoes an exponential decay to the rest state, here we find a damped oscillatory decay to the rest state. Such a damped oscillation is again reminiscent of the resonate-and-fire neurons. To reiterate, the oscillatory dynamics results in resonate-and-fire neurons from the state variable being complex. Further discussion of resonate-and-fire neuron characteristics of the nonlinear microring will be presented later.  

  \begin{figure}[h!]
        \centering
        \includegraphics[width=0.3\textwidth]{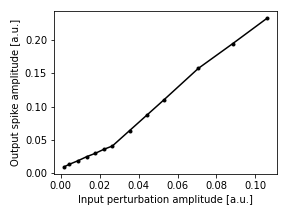}
        \caption{Output spike amplitude as a function of input perturbation amplitude illustrating a threshold behavior.}
        \label{fig:thres}
    \end{figure}
    
The presence of a spiking threshold is evidenced in Fig. \ref{fig:thres}, which shows the output spike amplitude ($P_{out}$) as a function of the input perturbation amplitude $P_{in}$. Unlike typical integrate-and-fire spiking neurons with a rectified linear unit (ReLU) like transfer function, the transfer function here has more of a leaky ReLU transfer function i.e. $P_{out} \neq 0$ for $P_{in}$ below the threshold.

\begin{figure}[h!]
    \centering
    \includegraphics[width=0.35\textwidth]{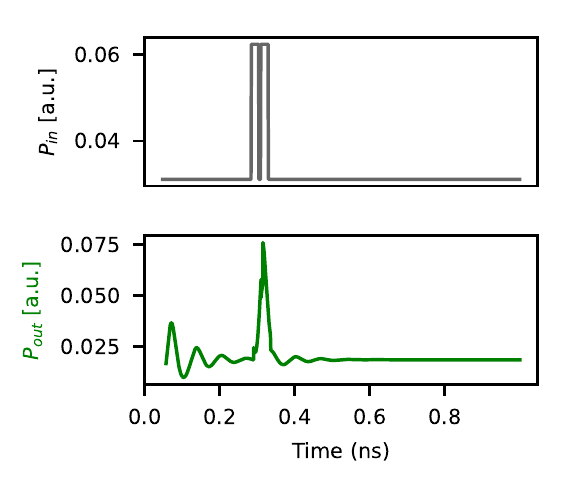}
    \caption{Temporal integration. (top) Input pulse train, with closely spaced pulses, (bottom) Corresponding output showing a single spike. }
    \label{fig:temp}
\end{figure}

Another critical feature of a spiking neuron is temporal integration, which means the device response is proportional to the power integral within some temporal window. To check for this feature, we study the device response to a doublet of closely spaced input pulses, each of whose amplitude is below the spiking threshold, shown in Fig. \ref{fig:temp}. The corresponding output shows a single spike in response to the input pulse doublet; this  means the response is dependent on the total energy i.e. the integral of power.   

\begin{figure}[h!]
    \centering
    \includegraphics[width=0.35\textwidth]{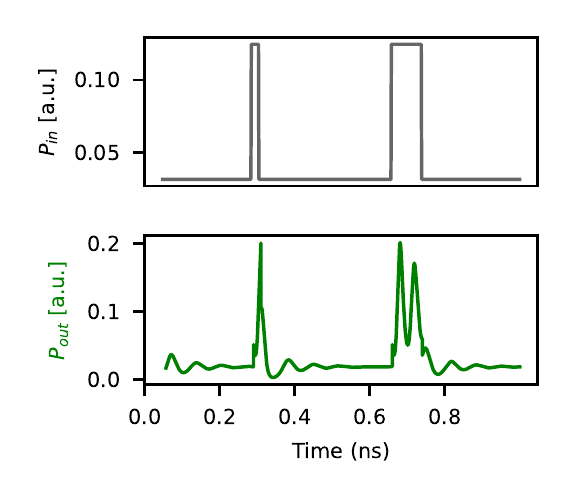}
    \caption{Pulse energy encoding: (top) input pulse sequence with different pulse widths, (bottom) corresponding output spikes where the input pulse energy information is encoded into the number of output spikes.}
    \label{fig:pulse}
\end{figure}

While the property of temporal integration reveals that the microring encodes the energy within the input pulses, we further study the second-order property of pulse-energy encoding in the microring. We first characterize the system response to input pulses of different pulse widths shown in Fig. \ref{fig:pulse}.  In the short input pulse width limit (left pulse), the system responds with a singular spike. On the other hand, in the long pulse width limit (right pulse), the system generates a pulse doublet. We find that increasing the pulse width further results in bursts of spikes. With identical input peak power, the pulse width is a measure of the pulse energy; these results reveal the pulse energy encoding feature of the spiking dynamical system. Additionally, the ability of the system to generate spike bursts, where the inter-spike timing encodes input pulse information, is valuable for reliable synaptic transmission between neurons as well as selective activation \cite{izhikevich2003bursts}. \\

\begin{figure}[h!]
    \includegraphics[width=0.5\textwidth]{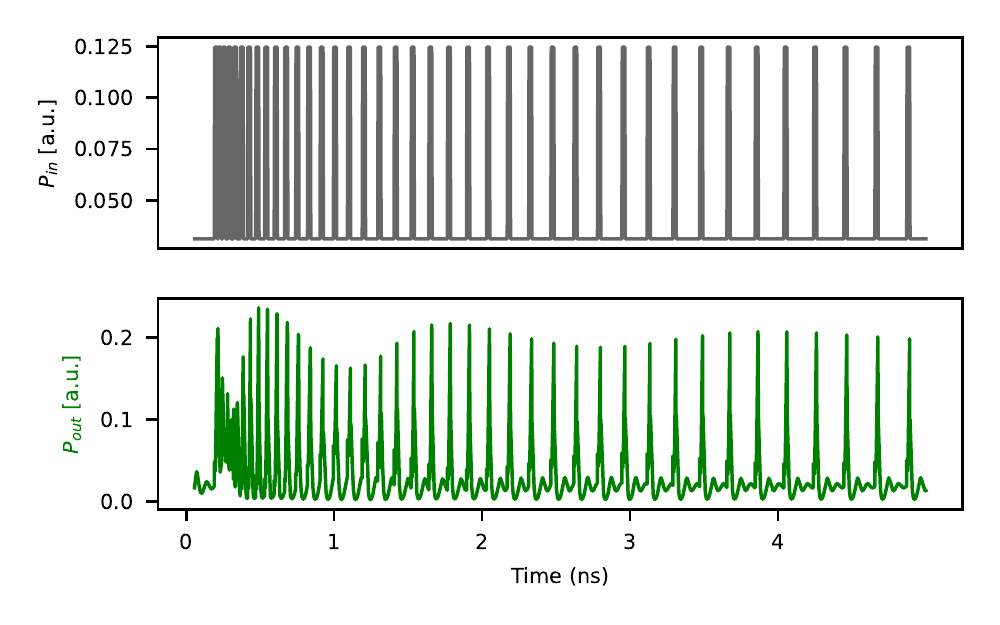}
    \caption{Input frequency dependent spike processing. (top) Input pulse train, with linearly increasing spacing between adjacent pulses, (bottom) Corresponding output spike train, showing a periodic dependence of output spike amplitude on input pulse frequency . }
    \label{fig:eigen}
\end{figure}

To further verify the \textit{resonate-and-fire} neuron characteristics of the microring, we investigate its response to the frequency of input pulses. In \cite{izhikevich2001resonate}, such neurons are said to have an intrinsic eigenfrequency and are thus sensitive to the frequency of the input pulses. We simulate the device response to an input pulse train with linearly increasing inter-pulse spacing. Fig. \ref{fig:eigen} reveals a periodicity in the output spike amplitude as a function of inter-pulse spacing. The oscillation decays in the long inter-pulse spacing regime. This periodicity suggests that the neuron response has some dependency on the input frequency which is an expected characteristic of this class of neurons.  

\begin{figure}[h!]
    \centering
    \includegraphics[width=0.4\textwidth]{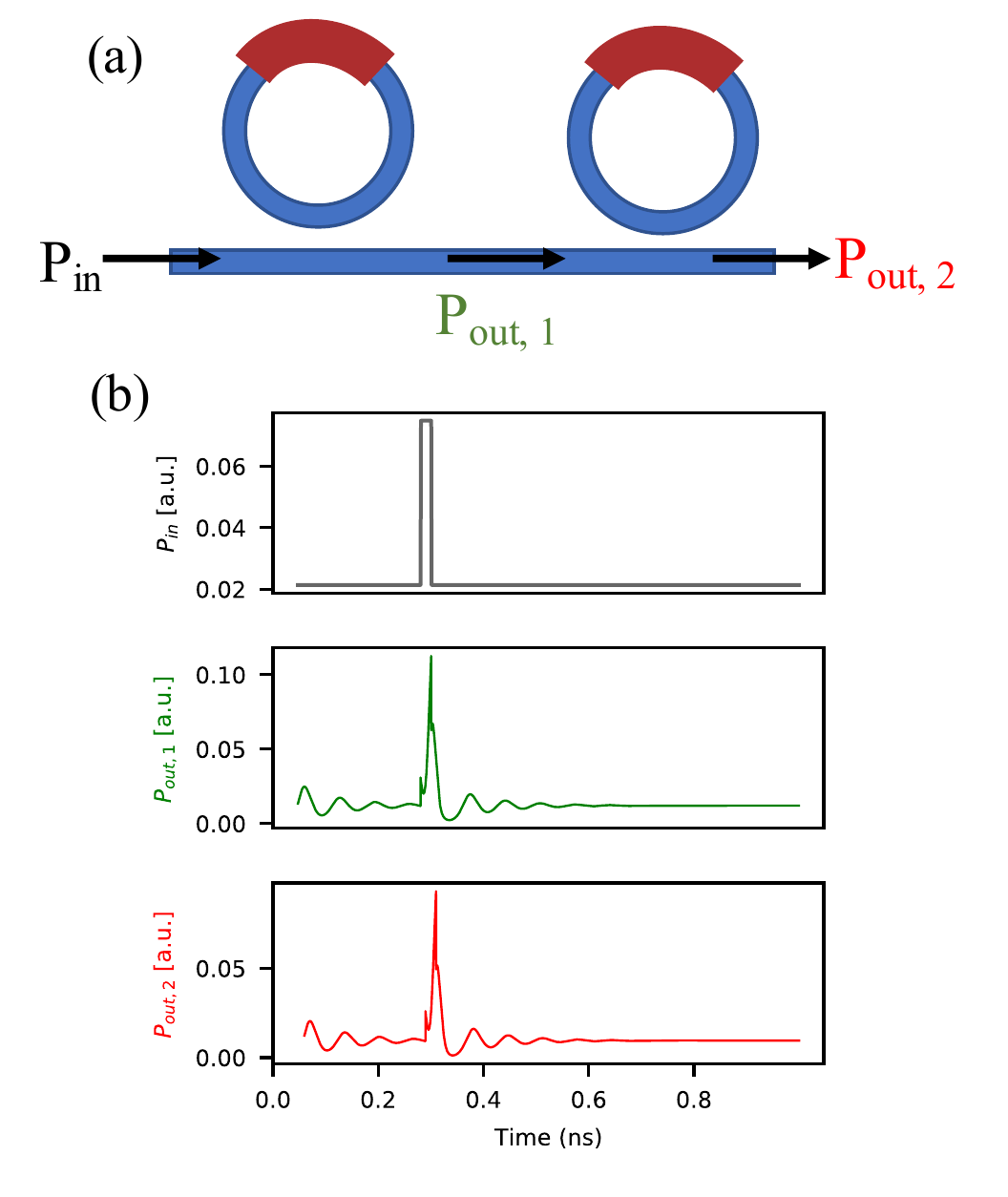}
    \caption{Cascadability in a two-neuron feedforward circuit.(a) Schematic illustration of two spiking microrings connected in a feed-forward fashion, where the output of the first MRR, $P_{\text{out,1}}$, drives the second MRR. (b)Demonstration of cascadability:  (top) input perturbation to the first MRR, (middle) output response of the first MRR, (bottom) output response of the second MRR to the spike output of the first MRR. }
        \label{fig:cascade}
\end{figure}

To verify the feasibility of using the microring as a spiking neuron, it is also important to check for cascadability, which is the ability of a neuron to excite a subsequent neuron. A simple demonstration of this is to feed the neuron output into an identical neuron, as shown in Fig. \ref{fig:cascade} and observe its response. Fig. \ref{fig:cascade} shows the spike output, labelled as $P_{\text{out,1}}$, in response to an input perturbation, $P_{\text{in}}$. This output spike is then fed into an identical neuron as a perturbed input. The response to this spike by the second neuron is shown, labelled as $P_{\text{out,2}}$. This demonstrates that the output spike of a neuron is capable of exciting a subsequent neuron. Given that both the neurons were identically biased in this simulation, this can also effectively serve as a demonstration of an autapse circuit, which is a feedback circuit where the output is fed back as the input. We thus demonstrate cascadability in the proposed spiking neuron. 

\section{Comparison of photonic spiking neurons}

In this section, we will delve into the merits and drawbacks of various spiking neuron devices in the context of implementation in a large network. Of course the overall performance of the network will depend on factors outside the neuron. Network architecture determines the number of interconnecting synapses, which scale quadratically with the number of neurons (in a fully-connected configuration). So for a large network, costs associated with synapses may overwhelm the overall network costs. However, here we limit our focus to spiking neurons themselves. To this goal, we have identified relevant performance metrics, such as power consumption, neuron firing rate i.e. processing speed, latency and footprint. We compare a representative subset of optoelectronic and all-optical devices against these metrics as enlisted in Table II.

\begin{table}[h!]
\caption{Performance metrics of state-of-the-art photonic spiking neurons.}
\resizebox{\columnwidth}{!}{\begin{tabular}{llllll}
\textbf{Device}           & \textbf{Platform} & \textbf{Power [mW]} & \textbf{Firing rate [GHz]}            & \textbf{Footprint [$\mu$m$^2$] }                       \\ \hline
DFB Laser \cite{nahmias2020laser} & InP      & 260           & $\sim$2                           & \textgreater{}600x200  \\
Graphene laser \cite{shastri2016spike} & InGaAsP-C-Si & 88 & 4 & 40  \\
Microdisk Laser \cite{alexander2013excitability} & InAsP-Si & .001 & 1  & 25$\pi$  \\ 
RT-PD \cite{romeira2013excitability} & InGaAsP & 3 & .002 & 600 \\ \hline
VCSEL \cite{hurtado2012investigation}, \cite{hurtado2010nonlinear}     &   Discrete       &    0.1     &   ~1                                                     &               40,000                   \\
2D PhC   \cite{yacomotti2006fast}           & InP      & $\sim$3        & 0.005                       &  500                  \\
MRR     \cite{van2012cascadable}         & Si       & 4              & 0.005                         & 25$\pi$              \\
PCM-based cavity \cite{chakraborty2018toward} & SiN      & 30              & 0.02                                       & 3600$\pi$ \\
Graphene-Si MRR & Si & ~100 & 40  & 25$\pi$            \\ 
\end{tabular}}
\label{metrics}
\end{table}

First, we compare the devices on a simple scale of physical size. This can be the primary constraint in edge computing devices. Integrated devices are naturally better than discrete ones in this regard as can be seen in Table \ref{metrics}.  Within integrated devices, platform-specific properties like optical material loss and refractive index further limit the footprint. Passive silicon-based devices tend to be the most compact due to the high index of silicon. Incidentally, these devices host all-optical spiking neurons. Additionally, electrical components, like photodetectors, interfacing optoelectronic devices can further add to their footprint. 

The next consideration is firing rate, or alternatively, the processing speed of a spiking neuron. Firing rate determines the temporal resolution of the spike-based processing. High-speed ($\geq$ GHz) processing is where neuromorphic photonics can really outshine electronics, which makes processing speed a major metric to consider. All-optical devices relying on slow nonlinear phenomena like thermal effects, are the slowest ones. Excitable lasers can operate up to a few GHz timescales. For higher speed needs, passive silicon-based all-optical graphene based neuron proposed in this work stands out as the most promising candidate. 

Finally, we compare the power consumption associated with the devices enlisted here. One could naively expect that optoelectronic devices have significantly lower power usage than all-optical devices, considering the high-power requirements of nonlinear optical processes. However, once the cost of opto-electronic conversion is factored in, the difference is not as dramatic as one would expect. Among the devices shown in Table \ref{metrics}, optoelectronic laser-based devices have higher power consumption on average than all-optical ones. In principle, the power consumption of cavity-based devices can further be reduced by implementing photonic cavities with smaller mode volumes like photonic crystals.  

Overall it is fair to say that there is no ideal spiking device that beats everything else across all metrics. There are additional challenges in both optoelectronic as well as all-optical approaches to really portray the full picture. For optoelectronic devices, we did not consider the speed limits due to parasitics in electronic links that can bottleneck the neuron processing speed. This may be addressed by heterogeneous integration with a CMOS chip via flip-chip bonding \cite{thacker2010flip} allowing for use for high-speed CMOS electronic components with a low parasitic pathway. Similarly, in all-optical devices, optical power cascadability is a concern, as optical power is inevitably lost at each stage of the network. In the future, waveguide amplifiers \cite{kik1998erbium} may alleviate these concerns by allowing for amplification every few stages.

\section{Training spiking neural networks}
Training refers to adjusting synaptic weights in the process of optimizing for a cost function, which can correspond to the difference between the target and actual output of a neural network. In the context of artificial neural networks, neurons represent static, continuous-valued nonlinear activations that are differentiable, which means they can be trained using gradient-descent based supervised learning algorithms like error backpropagation \cite{lee2016training}. In contrast, spike-based processors operate on dynamic temporal data that are an accumulation of delta function-like spikes that are non-differentiable, hence conventional backpropagation algorithms don't apply. Despite the touted benefits in energy-efficiency and noise robustness of SNNs relative to ANNs, the performance of SNNs on most machine learning tasks lag behind the ANNs primarily due to the issue of trainability \cite{lee2016training, roy2019towards}. This is further aggravated by the scarcity of dynamic dataset compatible to SNNs in contrast to the availability of large labelled static datasets for ANNs.        
Of course this impediment to SNN has attracted a lot of interest from neuroscientists and computer scientists alike to develop training algorithms catered to spiking neural networks. These algorithms can broadly classified into two categories: 
\begin{itemize}
    \item \textit{Conversion-based algorithms:} this approach aims to leverage backpropagation methods in ANNs by training the network as an ANN and then converting to an SNN \cite{7280696}, \cite{sengupta2019going}. Essentially, the static input (eg. pixel intensity in an image classification task) is converted to a spike train based on encoding schemes, such as rate encoding \cite{6497055}. An activation function that functionally resembles that of a spiking neuron, such as a rectified linear unit, ReLU, is chosen. Once training is done, weights are normalized to avoid arbitrarily high firing rates and the spike output is converted back to a static, analog value using the complementary decoding scheme. The merit of this approach is that it circumvents the difficulties of training temporal signals and can use existing frameworks like TensorFlow~\cite{Abadi2016TensorFlowAS} and Pytorch~\cite{Paszke2019PyTorchAI}. However, it suffers from the vanishing forward spike propagation problem where firing rates progressively decrease through network layers. Also, you pay for the training convenience in longer inference times owing to the conversion \cite{roy2019towards}. This approach has been considererd in photonic systems simulations -- Ref. \cite{chakraborty2019photonic} show an MNIST image classification using a rate encoding approach. Ref. \cite{hejda2021neuromorphic} even showed online rate encoding using a VCSEL based neuron.

    \item \textit{Spike-based algorithms:} this approach attempts to train using temporal data, without any conversion, and thus benefits from the sparsity and efficiency of spiking neural networks. Neuroscientists have identified various learning rules, based on neuroplasticity in actual biological synapses, known as spike-timing dependent plasticity (STDP). In STDP, weight is adjusted in response to the relative time difference between spikes of the connecting neurons. It has been used for both supervised and unsupervised learning. The spatial and temporal locality of STDP endows network-level effects such as reduced latency. The early works used STDP to perform spike based learning in a supervised manner, examples include ReSuMe \cite{ponulak2010supervised} and Tempotron \cite{gutig2006tempotron}. STDP has been abundantly demonstrated in photonic systems, starting out in \cite{fok2013pulse} to more recent implementations in VCSEL-based systems\cite{9018042}, \cite{8693533}, \cite{9142407}, \cite{feldmann2019all}. More recent supervised algorithms mirror backpropagation by approximating a differentiable function as the spiking nonlinearity and using gradient-descent based algorithms to optimize for the target spike train. Examples include SpikeProp \cite{bohte2000spikeprop}, NormAD \cite{Anwani2015NormADN}. There have been no implementations of such algorithms in photonic systems to the best of our knowledge. Typically, in software, unsupervised algorithms tend to have inferior performance over conversion-based supervised learning.   
  
\end{itemize}





\subsection{Benchmark MNIST handwritten digit classification simulation}
\begin{figure}[htbp]
    \centering
    \includegraphics[width=0.5\textwidth]{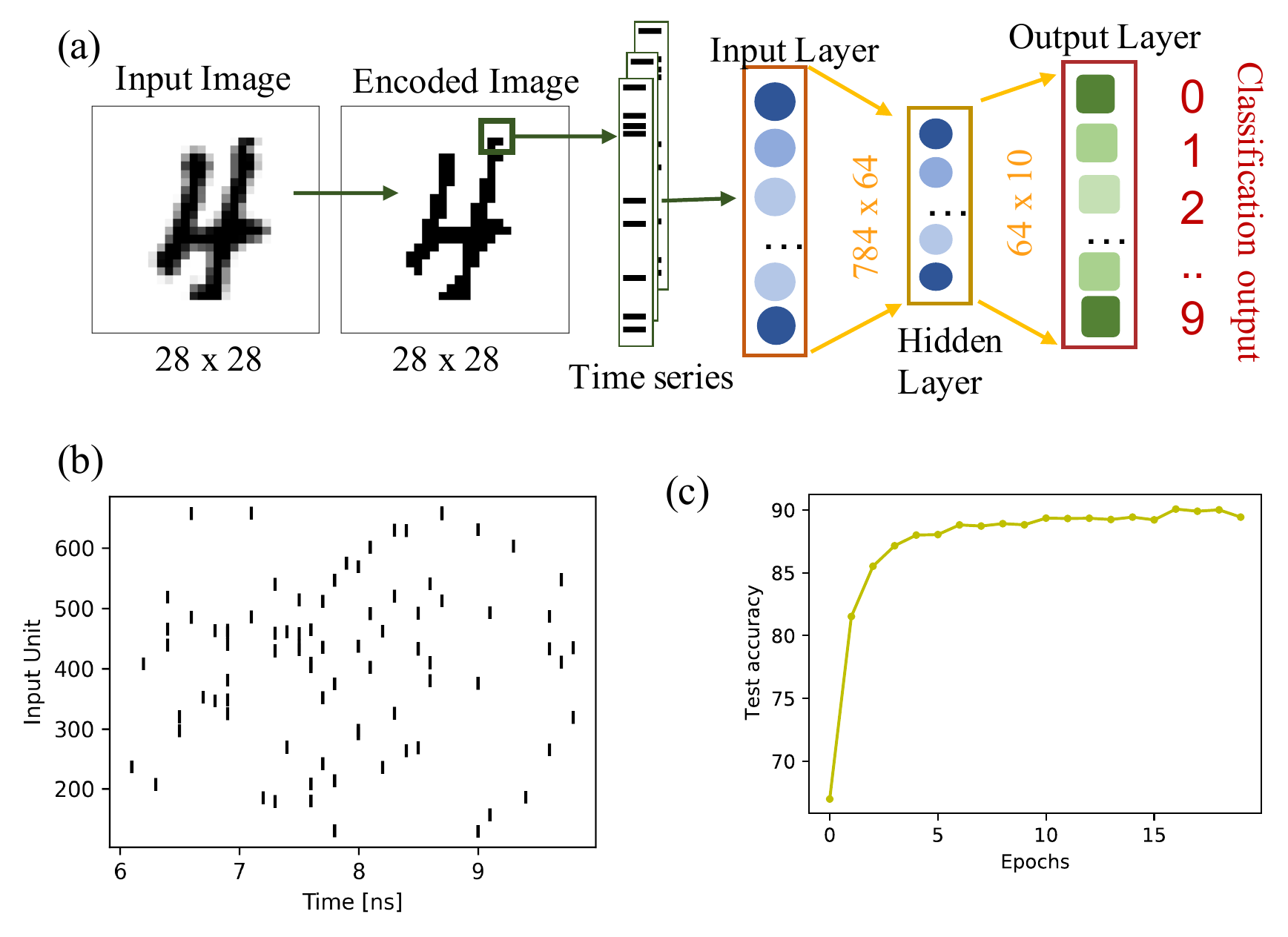}
    \caption{Benchmark MNIST handwritten digit classification simulation using spiking nonlinear nodes. (a) Schematic of the simulated classification network, showing a sample MNIST dataset image being encoded into a time series, and fed into a 3 layer fully-connected network. (b) Time-series equivalence of the encoded image in (a): each row corresponds to the time-series for a given pixel in the input image, (c) Inference classification accuracy as a function of epoch count. }
        \label{mnist}
\end{figure}

Here we employ a conversion-based algorithm to do a system-level analysis of our proposed spiking neuron. We simulate a three-layer fully-connected spiking neural network and study its accuracy in a benchmark MNIST handwritten digit classification task. The scheme is similar to typical artificial neural networks except the nonlinear units in the input and hidden layers are the spiking neuron units, and the inputs to the network are spike-based time series instead of analog values. This simulation is done using the Bindsnet simulation package \cite{10.3389/fninf.2018.00089} which allows simulating spiking neural networks within the Pytorch framework. The schematic of the network is shown in Fig. \ref{mnist}(a). Each input image in the MNIST dataset is composed of 28x28 pixels. First, each pixel in a given image is encoded using the Poisson encoding transform in Bindsnet and converted to a spike-based time series i.e. a format comprehensible by the spiking neurons. Fig. \ref{mnist}(b) shows the time series representation of the 784 pixels on the sample image. The frequency of the time series is proportional to the pixel intensity. \\
Each time series is then fed into an input layer with 784 (=28x28) neuron nodes. To emulate the spiking nonlinearity in each node, we employ the transfer function of the spiking neuron (shown in Fig. \ref{fig:thres}) as the nonlinear activation function. The input layer is fully connected to a hidden layer with 64 nodes, resulting in 784x64 = 50176 trainable parameters. The hidden layer is subsequently fully-connected to an output layer with 10 nodes, corresponding to the 10 possible digit label outputs, resulting in additional 64x10 = 640 trainable parameters. We use the log softmax function as the nonlinearity of the nodes in the output layer, which is typical for such multi-label classification networks.  

At each time step, the input value for a given pixel is either a 1 or a 0 corresponding to the presence or absence of a spike event. The input time series propagates through the network, and results in an output neuron in the output layer to spike at the highest rate for a given image. Then this neuron is compared to the label associated with the image and a negative log likelihood loss is calculated. While training the network, a stochastic gradient descent-based optimizer (Adam) was used to minimize the loss. 

In the inference stage, accuracy is computed by calculating the percentage of images that were correctly classified by the network. Fig. \ref{mnist}(c) shows the inference accuracy as a function of epochs, where the network is shown to converge to $93\%$ accuracy in about 10 epochs. This result demonstrates the feasibility of using the proposed spiking neuron in a spiking neural network. Our simulated network is much smaller than the one proposed in a similar work \cite{chakraborty2019photonic}, where 500 neurons were used in the hidden layer, although we achieved a slightly lower accuracy. 

\subsection{Training in photonics}
As discussed earlier, training spiking neural networks in itself remains an unresolved problem. The challenges compound when we translate the problem into the context of photonics, where process variations and operational constraints can further aggravate performance. Nevertheless, photonics can benefit from advances in algorithm research within computer science or electronics community, but it will be critical to judge which approach translates nicely to photonics. Here we present our outlook on the approaches to training photonic spiking neural networks.          

Conversion-based training algorithms based on rate-encoding strategies have had better performance over spike-based ones for conventionally ANN-oriented tasks like classification. Due to the convenience of this approach and the availability of dataset for classification tasks, this will likely be the popular approach for researchers to demonstrate the functionality of their spiking devices in the near term. However, this approach only allows for offline training and online inference. And we are yet to see how this approach will pan out when actually implemented on a photonic hardware. Robust characterization of photonic devices can be expected to be incredibly difficult, which means the discrepancy due to variations in hardware operation will cause performance degradation during online inference. Additionally, the conversion steps will add to the inference time, which may eliminate applicability in real-time applications.  

The ultimate way to ensure error-resilience to process variation and noise in photonics is online learning. Such approach can allow for algorithm-hardware co-design such that hardware-induced errors can be reliably mitigated. As discussed before, spiking hardware is not compatible with global learning rules like back propagation. Instead, to enable online learning in spiking hardware will require spatially local learning rules. Besides online learning, local learning allows leveraging the full potential of spiking networks, in terms of low inference latency. A common local learning rule in spiking, as we've discussed before, is STDP. STDP in photonic hardware have exclusively been shown in active platforms, using cross-gain modulation in SOAs \cite{toole2015photonic}, \cite{ren2015optical}, vertical cavity SOAs (VCSOAs) \cite{8693533}, \cite{8522021}. As discussed earlier, for photonic spiking to be a viable technology, compatibility to CMOS is crucial. The singular proposal of implementing STDP on a passive photonics platform was in Ref. \cite{9083359} using nonlinear optical effects. Ref.\cite{feldmann2019all} used an STDP-like learning approach through a feedback loop mechanism. While STDP is amenable to local learning, it does not really optimize towards a global objective function, and has had lower accuracy when compared to backpropagation methods \cite{falez2019unsupervised}. Higher performance may be obtained by using local learning rules that can allow for some form of global optimization through local weight update rules, akin to the Hopfield network that always seeks to minimize the global energy function. We thus foresee algorithms with local learning with a global optimization scheme to be the most suitable route when it comes to training photonic spiking hardware.


\section{Application domain of photonic SNNs}

Most application-oriented works in spiking neural networks have resorted to classification tasks as a means to compete against artificial neural networks. However, due to the lack of adequate training algorithms for SNNs, they perform relatively poorly in terms of accuracy. SNNs are underutilized if limited to such classification cases; their temporal encoding feature is better suited to applications with sparse dynamic data. Sample applications include natural language processing, event-based sensing and processing (high-speed navigation and localization: object tracking, scene reconstruction), etc. which would be prohibitive to carry out in conventional ANNs in an energy-efficient manner.

Event-based sensors output data in the form of spikes and spike-based processing naturally fits into the task of processing signals from such sensors. Applications within event-based sensing span tracking, robotics, object tracking, etc \cite{pfeiffer2018deep}. However, implementing application-oriented spike-based processing on a spiking hardware remains largely outstanding, except for a few works -- Ref. \cite{8100264} showed the use of TrueNorth, a spiking electronic hardware, for real-time gesture recognition from a dynamic vision sensor. There have not been any demonstrations of such applications in a spiking photonic hardware yet. The low-latency, high energy efficiency and processing speed of photonics makes it even more suitable for such applications. For instance, autonomous driving simultaneously requires low power and high speed processing, where conventional ANNs have significant computational overhead and latency \cite{wang2020temporal}. Another application can be brain-machine interfaces since SNNs can process biological spikes without transformation and offer low energy consumption, low latency, low thermal dissipation as required by these systems \cite{1300781}. Finally, the simultaneous high-speed and low power operation of spiking neurons can be beneficial for high-speed radio-frequency (RF) signal processing. Cognitive radio networks require high-speed decision making for resource allocation, and it has been shown in simulation that the spatio-temporal dynamics of spiking networks can enable such functionality \cite{8435955}.  

Even with classification-like tasks, SNNs may enable continual learning that ANNs are incapable of \cite{roy2019towards}. In deep learning models based on continuous-valued neurons, the network adapts to new datasets, and forgets old patterns in the process. This is in stark contrast to the human brain where memory tends to be permanent, barring any injuries or disease. With the additional temporal dimension in SNNs, SNNs may potentially enable continual learning \cite{roy2019towards}. Applications beyond the reach of conventional ANNs and spiking electronics are the real opportunities where photonic spiking hardware can fully demonstrate its computational might.    


\section{Conclusion}

Spiking neural networks offer the possibility of computationally powerful, noise-resilient next-generation neuromorphic processors. However, conventional computers are incompetent for such a distributed processing model which advocates for a specialized hardware. Electronics faces fundamental bottlenecks in interconnectivity and bandwidth, and presents an opportunity to capitalize on photonics. Innovations in developing photonic spiking hardware has lasted just over a decade, and we may now be at the cusp of enabling a real technological platform capable of real applications.       

The landscape of photonic spiking neurons can broadly be classified into optoelectronic devices including excitable lasers and all-optical devices, including nonlinear photonic cavities. Laser-based devices on discrete or active platforms may not be as conducive to scalability as the CMOS-compatible devices. To this end, we have proposed a novel spiking neuron design based on a graphene-on-silicon microring resonator. We have presented preliminary simulation results showing spiking neuron-like behavior of the microring device, which arises from the interplay of nonlinear optical effects in both graphene and silicon. 

As we widen the lens from devices to systems based on photonic spiking hardware, we will need to consider algorithms that can optimize the performance of the network for a given application while ensuring resiliency to process variations and noise in hardware. Local online learning rules can be a good avenue for exploration that may be able to accommodate for the needs of photonic hardware.                                                              
Finally, it is important to carve out the application space of photonic spiking neuron technology -- likely outside the scope of ANNs and spiking electronics. These might include event-based sensing applications, autonomous control, etc. Only once photonic spiking neurons can enable applications beyond what was previously possible with spiking electronics or ANNs will they be worth the research efforts spent on them. The confluence of innovations in spiking neuron devices and in algorithms paints a hopeful future for what is ahead.

\section*{Acknowledgment}
The authors would like to thank Office of Naval Research (ONR) for the funding support.

\bibliographystyle{IEEEtran}
\bibliography{bibliography}

\end{document}